\documentclass[11pt]{article}

\usepackage{multicol,multirow}
\usepackage{booktabs, colortbl}
\usepackage{longtable}
\usepackage{tabularx}
\usepackage{graphicx}
\usepackage{colortbl}
\usepackage{amsmath}
\usepackage{booktabs}
\usepackage{amsthm,amsopn,amsxtra,latexsym,dsfont} 
\usepackage{bm}      
\usepackage{amssymb}
\usepackage{listings}
\usepackage[table]{xcolor}
\usepackage{float}
\usepackage{xcolor}
\usepackage{dsfont}

\usepackage{ragged2e}
\usepackage[figuresright]{rotating}
\usepackage[font=small,skip=0pt]{caption}
\usepackage{indentfirst}

\usepackage[sc]{mathpazo}
\usepackage{fullpage}
\usepackage[authoryear,sectionbib,sort]{natbib}
\usepackage{graphicx}
\usepackage{adjustbox}
\usepackage[boxruled,linesnumbered]{algorithm2e}
\usepackage{multirow}
\usepackage{amsmath}
\usepackage[utf8]{inputenc}
\usepackage{lineno}
\usepackage{titlesec}
\usepackage{graphicx}
\usepackage{subcaption}
\usepackage{booktabs,multirow}
\usepackage{here}
\usepackage{url}

\titleformat{\section}[block]{\Large\bfseries\filcenter}{\thesection}{1em}{}
\titleformat{\subsection}[block]{\Large\itshape\filcenter}{\thesubsection}{1em}{}
\titleformat{\subsubsection}[block]{\large\itshape}{\thesubsubsection}{1em}{}
\titleformat{\paragraph}[runin]{\itshape}{\theparagraph}{1em}{}[. ]

\title{Unified Multivariate Ordinal Model for analysis of sensory attributes}

\author{Janaína Marques e Melo$^{1}$ \and 
        João César Reis Alves$^{1,\ast}$ \and
        Gabriel Rodrigues Palma$^{2}$ \and 
        Sílvia Maria de Freitas$^{3}$ \and
        Idemauro Antonio Rodrigues de Lara$^{1}$}

\date{}

\begin{document}

\maketitle

\noindent{} 1. University of São Paulo, Piracicaba Brazil;

\noindent{} 2. Maynooth University, Maynooth, Ireland;

\noindent{} 3. Federal University of Ceará, Ceará, Brazil;

\noindent{} $\ast$ Corresponding author; e-mail: joaocesar@up.br

\bigskip


\bigskip

\textit{Keywords}: Balanced incomplete blocks; Hedonic scale;  Maximum likelihood; Prebiotic beverage.

\bigskip

\textit{Manuscript type}: Research paper. 

\bigskip

\noindent{\footnotesize Prepared using the suggested \LaTeX{} template for \textit{Am.\ Nat.}}

\newpage{}

\section*{Abstract}
Experiments involving sensory analysis of foods and beverages are beneficial for selecting healthy products and assessing the preferences of potential consumers. They are generally planned in incomplete blocks, and their attributes, such as aroma, colour, and flavour, are evaluated using a 9-point hedonic scale, characterizing an ordinal variable response. Also, the generalised logit model with random effects for panellists is one of the appropriate models to relate the multivariate response to the covariates. This study aims to present a method for analysing sensory attributes through a unified multivariate model. Due to the nature of the variable, each separate model already corresponds to a multivariate analysis, so our proposal would incorporate a complete analysis with solely one model. This proposal is based on multivariate methods for categorical data and maximum likelihood theory. Our method was evaluated through a simulation study, in which we consider three distinct formulations with two attributes to represent various formulation selection scenarios via mixed discrete models. The simulated results demonstrated overall concordance rates exceeding 80\% for the unified model compared to the separate models. Moreover, as motivation is presented a study of 13 prebiotic beverages based on cashew nut almonds added to grape juice,  with 130 potential consumers. The attributes evaluated were overall impression, aroma, Body, sweetness and flavour, using a 9-point hedonic scale. The selected unified model considering all attributes was the non-proportional odds mixed-effect model. According to this model, the prebiotic beverage formulations most likely to be accepted were: 8\% sugar and 40\% grape juice ($F_4$), 6\% sugar and 44\% grape juice ($F_6$), and 9\% sugar and 30\% grape juice ($F_{13}$). The results obtained by this approach were according to the analyses for each attribute. However, the unified analysis and computational time showed advantages of this proposal.

\newpage{}

\section{Introduction}
There is a growing interest from consumers for foods that, in addition to the basic function of nourishing, promote beneficial health effects \citep{Prates2002, Franco2006}. This new consumption profile has stimulated industries to research and develop products that meet these criteria. A sensory analysis survey is usually carried out to assess consumer acceptance of certain products, in which attributes (or characteristics) are classified according to a previously established response scale based on ordinal categories, known as the hedonic scale. In the hedonic scale test, product acceptability is measured through a hierarchy based on ordinal categories that measure different degrees of liking \citep{Dutcosky2011}. \citep{Nguyen2019} used this technique to compare sensory attributes in food products.

Through sensory acceptance tests, the judges can give their opinion on different product characteristics related to appearance, aroma, flavour and texture using the sense organs (smell, touch, taste, sight and hearing). In this way, there is a sensorial analysis because it deals with the organs of the human sense. According to \citep{Chaves1998}, in Brazil, sensory analysis began in 1954 with tasters for the classification of Brazilian coffee. Its use extends from monitoring, improving or launching new products on the market to analyzing the effect of packaging \citep{Rebouccas2016, Borem2019}. In developing new products, evaluating consumers' responses in the initial stages and not only in the final product is of fundamental importance. Therefore, it is necessary to optimize the formulation of the product to evaluate the different concentrations of the ingredients used, which can lead to its acceptance. In this way, it is possible to obtain the formulation that allows reaching the greatest possible acceptability \citep{Rebouccas2016b, Rebouccas2018}. 

According to \citep{Dutcosky2011}, sensory evaluation provides technical support for research, industrialization, marketing and quality control. In this way, the importance of applying sensory analysis is how much it influences the development of a particular product. The sensory quality of the food and its maintenance favour consumer loyalty to a specific product in an increasingly demanding market \citep{Teixeira2009}. Thus, a product that is developed based on prior knowledge of sensory attributes, as well as having a precise control of this sensory quality, generally becomes more attractive, bringing an advantage to the company by producing a product that meets the new desires of the consumer.

In some sensory research, the number of samples to be evaluated can be numerous, making it impossible to form complete homogeneous blocks. There may also be an interest in using a small number of products per individual for economic or logistical reasons or the nature of the experiment. In addition, this product diversity can compromise the indication of the best, and there is a possibility of promoting fatigue among the participants. In this context, an experiment in incomplete blocks is suggested, in which the tasters do not taste all the product options under test. According to \citep{montgomery2013design}, situations like this usually occur due to experimental setups or the physical size of the block. Also, the author reported that when all treatment comparisons are equally important, the treatment combinations used in each block should be selected in a balanced way so that any pair of treatments occur together the same number of times as any other pair. The following relationships are necessary conditions to have a balanced incomplete block design: $rt = hb$, $\lambda(t - 1) = r(h - 1)$,
$r > \lambda$, $b \geq t$.

Assume that there are $t$ treatments and $b$ blocks, each block contains $h$ treatments, and each treatment occurs $r$ times in the experiment (or is replicated $r$ times). This methodology is widely used in the evaluation of wines, as was the case of the study \citep{Idolo2019} that used a balanced incomplete block design to evaluate the effect of ageing temperature on \textit{hibiscus sabdariffa} (roselle) wine. 

Moreover, these studies use categorical variables (hedonic scales) to assess sensory attributes analyzed in isolation by attribute. Examples of individual analyses by attribute include, among others: \citep{Li2014} to select coffee flavours;  \citep{Lemos2015} to evaluate cashew apple nectar brands; \citep{Aaby2019} that determined the relationships between the sensory attributes of eight genotypes of red raspberry fruits (\textit{Rubus idaeus L.}); \citep{Fatoretto2018} to choose two types of tomatoes using flavour; \citep{Rebouccas2018} and \citep{marques2023prebiotics} to selection of prebiotic beverage formulations.  

This article aims to present a unified model for analysing various sensory attributes together, using a unified multivariate model, including random effects to consider the structure of the experimental design. Additionally, this work is motivated by a sensory study with grape juice developed by \citep{Rebouccas2016}. In this way, we also hope this work can help potential sensitometry researchers by using a set of techniques that assist in decision-making regarding product selection.

\section{Method}
\subsection{Review on specific model for an attribute}

Let $\bm{Y}$ be the vector representing the response variable on the hedonic scale with $J$ categories related to a sensory attribute, where $\bm{Y} \sim Multinominal(\pi_1, \cdots, \pi_J)$. Each response category is associated with a covariate, here denominated formulation ($\bm{F}_t$, with $t = 1,2, \cdots, T$). The interest is in obtaining the probabilities of occurrence of each response category $\pi_j$, with $j = 1,2, \cdots, J$, associated with each formulation.

To associate the covariate (formulation) with the panellist response categories, a possible parsimonious model is the proportional odds model, which is a model that uses the accumulated probabilities of response categories, in which we assume odds ratios proportionality. Let $\theta_j$ the cumulative probability of occurrence until $ j-$th category for a given formulation effect, $\bm{F}$, defined by
\begin{eqnarray*}
	\theta_j   = \pi_1 (\bm{F}) + \pi_2 (\bm{F}) + \cdots + \pi_{j-1} (\bm{F}) + \pi_j (\bm{F})   = \mathds{P}(Y \leq j \mid \bm{F}),
\end{eqnarray*}
for all $j = 1, 2, \cdots, J$. 

The proportional odds model adopted in this work quantifies, probabilistically, the consumer attitude, considering the presence of the effect of the covariates. In this work, the mixed proportional odds model is used for each sensory response attribute separately, represented by\\
{
	\begin{equation}\label{MLCMCPmetodo1}
	g(\bm{F}_{},\bm{z}) = \ln \left[ \frac{\theta_j(F)}{1-\theta_j(F)} \right] = {\alpha}_{j} + \bm{\beta_{}^{\top}} \bm{F}, \hspace{.8em}
	\end{equation}
}\\
\noindent where $ {\alpha}_{j} $ is the intercept of the $j$-th response category for a given sensory, with $j = 1, 2, \cdots, J-1$, $ \bm{\beta}^{\top} = (\beta_{1}, \beta_{2}, \cdots, \beta_{t}) $ the regression parameter vector common $t$-th formulation, with $t = 1, 2, \cdots, T$.
Parameter estimates are obtained by the maximum likelihood theory, using the interactive Newton-Raphson process \citep{Azzalini2017}.
In terms of accumulated probabilities, the mixed proportional odds model is defined by \\
{
\begin{equation}\label{Eq.:ProbabilidadeTheta}
\theta_j(\bm{F}_{},\bm{z}) = \frac{\exp{({\alpha}_{j} + \bm{\beta_{}^{\top}} \bm{F})}}{1 + \exp{({\alpha}_{j} + \bm{\beta_{}^{\top}} \bm{F})}}, \hspace{1em} j = 1, 2, \cdots, J-1.
\end{equation}
}
To use the mixed proportional odds model, it is necessary to verify the assumption of proportionality of the odds ratios. For this, we use the Likelihood-Ratio Test (LRT) where the null hypothesis is that ${\beta}_j = {\beta}$ and the alternative hypothesis is that ${\beta}_j \neq {\beta}$ with $j = 1, \cdots, J$. The test statistic is given by
\begin{equation}\label{TRV}
\Lambda = -2 \log \left( \frac{L_{H_0}}{L_{H_A}}\right) 
\end{equation}
where $ L_ {H_ {0}} $ is the likelihood function under the null hypothesis, assuming the proportional odds model and $ L_ {H_A} $ represents the likelihood function under the alternative hypothesis, or that is, assuming non-proportionality. Additional details on the odds model can be seen at \citep{Agresti2010}.

\subsection{Unified multivariate cumulative logit model approach} \label{a2Secao:modeloproposto}

Let $\bm{Y}=(\bm{Y}_{1},\bm{Y}_{2}, \ldots,  \bm{Y}_{L})^{T}$ be the vector representing the response variable for all attributes, each one in the hedonic scale with $J$ response categories. The answer related to $i-$th panellist for the $l-$th attribute is denoted by the vector $\bm{Y}_i = (Y_{i1l}, \cdots, Y_{ijl}) ^{\top}$, where $Y_{ijl}$ represent the indicator variables for the response categories, that is, $Y_{ijl} = 1$ if the $i-$th panellist, when evaluating the l-th attribute, opted for $j-$th response category; and $Y_{ijl} = 0$ otherwise, for $l=1,2,\ldots, L$. 

The proposed model is based on the cumulative probabilities of the response categories, in which odds ratios are assumed to be proportional or not. Let $\theta_{j} = \mathds{P}(Y \leq j \mid \bm{F}, \bm{A})$, $j = 1, 2, \cdots, J$, the probability occurrence up to the $j-$th response category for a given beverage formulation ($\bm{F}$) of a given sensory attribute ($\bm{A}$). In this model, both covariates, $\bm{F}, \bm{A}$ are represented by dummy variables. 
Given this, the proposed unified model is represented by:
	\begin{equation}\label{a2eq:modeloproposto}
	 \ln\left[ \frac{\theta_{j}(\bm{F}_{},\bm{A})}{1-\theta_{j}(\bm{F}_{},\bm{A})} \right] = {\alpha}_{j} + \bm{\beta}_j^{\top}  \bm{F}  + \bm{\delta}_j^{\top} \bm{A} , \hspace{.8em}
	\end{equation}
where ${\alpha}_{j}$ is the intercept of the $j-$th response category, $\bm{\beta}_j^{\top} = (\beta_{j1}, \beta_{j2} , \cdots, \beta_{jT})$ the vector of parameters associated with beverage formulations, $\bm{\delta}_j^{\top} = (\delta_{j1}, \delta_{j2}, \cdots , \delta_{jL})$ the vector of parameters associated with the $L$ sensory attributes.
If we assume proportional odds then in the model (\ref{a2eq:modeloproposto}), $\bm{\beta}_{j}=\bm{\beta}$  and $\bm{\delta}_{j}=\bm{\delta}$ for all $j=1,2, \ldots, J$, as demonstrated in the previous section by means of the equations \ref{MLCMCPmetodo1} and \ref{TRV}. If the proportionality hypothesis is rejected, then regression coefficients, $\bm{\beta}_j$ and $\bm{\delta}_j$, for both formulations and attributes vary with each response category.

We also used the LRT  to verify the effect of the covariates  (formulation and sensory attribute), in which the null is $\bm{\beta}_j = \bm{0}$ (i.e., for the $j$-th category, there is no difference among formulations) and  $\bm{\delta}_j = \bm{0 }$ (i.e., for the $j$-th category, there is no difference among sensory attributes). In general, sensory studies are incomplete block designs, and the panellist are not trained. It is essential to consider including a random effect for the panellist. Thus, the mixed unified multivariate model is described by:
\begin{equation}\label{a2eq:modelopropostomisto}
\ln \left[ \frac{\theta_{j}(\bm{F}, \bm{A} \mid u_{i})}{1-\theta_{j}(\bm{F}, \bm{A} \mid u_{i})} \right] = {\alpha}_{j} + \bm{\beta}_{j}^{\top} \bm{F} + \bm{\delta}_j^{\top} \bm{A} + u_{i}, \hspace{.8em}
\end{equation}
in which,  in addition to the model (\ref{a2eq:modeloproposto}), $u_{i}$ is the random effect associated to the $i-$th panellist, where $u_ {i} \sim N(0,\sigma^2_{u_{}})$.
We have used the profiled likelihood procedure to obtain the 95\% confidence interval for the standard deviation of the random effect coefficient and to verify if the $\sigma^2_{u_{}}=0$ is included or not in the interval, i.e., if it does not contain the value zero, there is a random effect for the panellist.
In terms of cumulative probabilities, the model (\ref{a2eq:modelopropostomisto}) is defined by:
\begin{equation}\label{Probab_mixed}
	\theta_{j}(\bm{F}_{},\bm{A} \mid u_{i}) = \frac{\exp{({\alpha}_{j} + \bm{\beta}_j^{\top}  \bm{F}  + \bm{\delta}_j^{\top} \bm{A}+u_{i})}}{1 + \exp{({\alpha}_{j} + \bm{\beta}_j^{\top}  \bm{F}  + \bm{\delta}_j^{\top} \bm{A}+u_{i})}}, 
\end{equation}
with  $j=2, \ldots, J$  and  $l=2, \ldots, L$.
The computational procedure was developed using the package \verb"ordinal" \citep{Christensen2015} for the cumulative logit model (unified multivariate mixed model), available in software \texttt{R} \citep{Team2020}. The script used to perform the analysis is available in the repository \texttt{https://github.com/Idemauro}.

\subsection{Simulation study} \label{a2Secao:simulationStudy}

A simulation study was conducted to evaluate the effectiveness of the unified multivariate ordinal model in the analysis of sensory data. The primary objective was to investigate the concordance between the unified model and the separate models under different experimental scenarios. To optimize the simulation process, three distinct formulations~($t=1,2,3$), $F_1, F_2, F_3$, and two sensory attributes~($l=1,2$), A and B,  were organized within a balanced complete block design. We also considered as a response an ordinal polytomous variable, with 5 points $(1< 2<3<4<5)$.

The simulated data were generated based on the parameters of a mixed cumulative logit model assuming proportional odds (Equation \ref{a2eq:modelopropostomisto}). Thirteen distinct scenarios were considered, each with 1,000 data replications, using two different sample sizes: $N = 90$ and $N = 300$ panellists (blocks).

The concordance rates between the unified model (Equation \ref{a2eq:modelopropostomisto}) and the individual models for each attribute (Models A and B) are presented in Table \ref{tabelasimulation}.

\begin{table}[H]
\centering
\caption{Concordance rates between the unified multivariate ordinal model (Equation \ref{a2eq:modelopropostomisto}) and the proportional odds logit model, evaluated separately for each sensory attribute (Model for Sensory Attribute A and Model for Sensory Attribute B), across thirteen different simulation scenarios, considering the three formulations $(F_{1}, F_{2}, F_{3})$ for sample sizes $N=90$ and $N=30$.}
\label{tabelasimulation}
\begin{adjustbox}{max width=\textwidth, max totalheight=0.9\textheight, keepaspectratio}
\begin{tabular}{lccc}
\hline
\multicolumn{4}{c}{$N=90$} \\ \cline{2-4}
Scenarios & Unified Model & Model for Attribute A & Model for Attribute B \\ \hline
$F_{3}<F_{1}<F_{2}$ & 99.9\% & 99.9\% & 100.0\% \\
$F_{1}<F_{3}<F_{2}$ & 100.0\% & 100.0\% & 100.0\% \\
$F_{2}=F_{3}<F_{1}$ & 90.0\% & 98.2\% & 96.9\% \\
$F_{2}<F_{3}<F_{1}$ & 90.0\% & 100.0\% & 100.0\% \\
$F_{3}<F_{2}<F_{1}$ & 84.5\% & 99.3\% & 98.8\% \\
$F_{1}=F_{2}=F_{3}$ & 84.5\% & 99.8\% & 98.3\% \\
$F_{1}=F_{2}<F_{3}$ & 95.8\% & 99.7\% & 97.8\% \\
$F_{1}<F_{2}=F_{3}$ & 95.6\% & 93.5\% & 95.0\% \\
$F_{1}<F_{2}<F_{3}$ & 99.9\% & 99.9\% & 100.0\% \\
$F_{2}<F_{1}=F_{3}$ & 95.4\% & 96.1\% & 94.8\% \\
$F_{2}<F_{1}<F_{3}$ & 100.0\% & 99.9\% & 99.8\% \\
$F_{3}=F_{1}<F_{2}$ & 93.2\% & 99.4\% & 98.1\% \\
$F_{3}<F_{1}=F_{2}$ & 96.8\% & 93.8\% & 94.0\% \\ 
\hline
\multicolumn{4}{c}{$N=300$}  
 \\  \cline{2-4}
$F_{3}<F_{1}<F_{2}$ & 99.9\% & 99.9\% & 99.9\% \\
$F_{1}<F_{3}<F_{2}$ & 99.9\% & 99.9\% & 99.9\% \\
$F_{2}=F_{3}<F_{1}$ & 84.3\% & 91.1\% & 89.7\% \\
$F_{2}<F_{3}<F_{1}$ & 99.9\% & 99.9\% & 99.9\% \\
$F_{3}<F_{2}<F_{1}$ & 99.9\% & 99.9\% & 99.9\% \\
$F_{1}=F_{2}=F_{3}$ & 77.3\% & 83.4\% & 87.1\% \\
$F_{1}=F_{2}<F_{3}$ & 84.4\% & 91.2\% & 88.8\% \\
$F_{1}<F_{2}=F_{3}$ & 95.2\% & 95.6\% & 95.3\% \\
$F_{1}<F_{2}<F_{3}$ & 99.9\% & 99.9\% & 99.9\% \\
$F_{2}<F_{1}=F_{3}$ & 95.8\% & 95.9\% & 94.8\% \\
$F_{2}<F_{1}<F_{3}$ & 99.9\% & 99.9\% & 99.9\% \\
$F_{3}=F_{1}<F_{2}$ & 81.6\% & 90.7\% & 88.2\% \\
$F_{3}<F_{1}=F_{2}$ & 95.5\% & 94.6\% & 96.2\% \\ \hline
\end{tabular}
\end{adjustbox}
\end{table}

In Table \ref{tabelasimulation}, the simulation results are presented based on hypothetical scenarios defined by hierarchical relationships among the formulations $F_{1}$, $F_{2}$, and $F_{3}$. Each scenario was generated by assuming different evaluation patterns of sensory attributes, simulated for sample sizes of $N=90$ and $N=300$. The formulations were ranked according to their assumed quality, creating scenarios where, for instance, $F_{3}<F_{1}<F_{2}$ implies that $F_{2}$ is hypothetically the best evaluated formulation, followed by $F_{1}$ and $F_{3}$. The concordance between the unified and separate models was then calculated for each scenario, considering sensory attributes evaluated individually and collectively.

In the $F_{3}<F_{1}<F_{2}$ scenario, the unified model achieved a concordance rate of 99.9\%, while the separate models for sensory attributes A and B reported concordance rates of 99.9\% and 100.0\%, respectively. This high level of concordance highlights the ability of the unified model to adequately represent the observed patterns, particularly in scenarios with clearly defined hierarchies among the formulations.

Overall, the concordance rates for the unified model exceed 90\% in scenarios with well-defined hierarchies ($F_{1}$, $F_{2}$, $F_{3}$). However, in scenarios where the formulations have the same effect ($F_{1}=F_{2}=F_{3}$), the concordance rates tend to be lower, reaching 84.5\% for $N=90$ and 77.3\% for $N=300$. This suggests that the unified model has greater difficulty distinguishing identical formulations, a limitation also observed in the separate models.

Finally, when comparing sample sizes ($N=90$ and $N=300$), it is noted that increasing the sample size does not necessarily lead to substantial improvements in concordance rates. This result suggests that, in many scenarios, the model’s performance is more influenced by the complexity of the evaluation patterns among the formulations than by the number of available observations.

\section{Motivational study}
\label{s:material}

The data that served as a motivation study in this work came from the experimental design of the Department of Food Technology at the Federal University of Ceará (UFC), Fortaleza, Brazil, in 2016. This study combined the nutritional, functional and sensory characteristics using chestnut of cashew, grape juice, sugar and prebiotic substances to develop a beverage. The water-soluble extract of processed and raw cashew nut kernels, two prebiotic substances, inulin (polymerization degree $\geq$ 10, Orafti GR) and oligofructose (2 - 8 monomers, Orafti P95) and commercial crystal sugar. To add flavour, concentrated grape juice was used (pH =$2.99; 15.2^{o}$Brix), which was defined through preliminary studies. Using the $2^2$ factorial rotational central composite design, with five repetitions at the central point, using combinations of juice percentages (16\%, 20\%, 30\%, 40\% and 44\%) and sugar (3 \%, 4\%, 6\%, 8\% and 9\%), 13 prebiotic beverages were formulated ($F_1$ to $F_{13}$), with $F_9$ to $F_{13}$ those of the central point. More details can be seen in \citep{Rebouccas2016}. The evaluation of sensory acceptance of prebiotic beverage formulations was performed only in one session, with 130 untrained panellists (blocks). The beverage samples were served in a sequential monadic way, following a balanced incomplete block design, in which each panellist evaluated $h=4$ out of the $t=13$ proposed beverage formulations each repeated $r=4\times10$ times.
Originally, the scores assigned to the drinks were based on the 9-point hedonic scale ("9 = I liked it extremely", "8 = I liked it very much", "7 = I liked it moderately", "6 = like slightly", "5 = neither like nor dislike", "4 = dislike slightly", "3 = dislike moderately", "2 = dislike very much", "1 = extremely disliked") in terms of acceptance of the sensory attributes of overall impression, aroma, Body, sweetness and flavour. However, in this work, due to the occurrence of sparse data and to avoid overparameterization, in this work, the original 9-point hedonic scale was reduced to 5 points, where "5 = I liked it extremely or very much", "4 = I liked it moderately or slightly", "3 = neither liked nor disliked", "2 = disliked slightly or moderately", "1 = disliked very or extremely".

\section{Data analysis and Discussion}

Initially, an exploratory analysis was carried out using a multiple correspondence analysis. Correspondence analysis is a multivariate technique beneficial for categorical data on an ordinal or nominal scale. It applies to the contingency table of observed probabilities, considering the response categories and the levels of an explanatory covariate, reducing it to a two-dimensional matrix by means of singular value decomposition, which can be represented graphically (two-dimensional plot) \citep{Johnson2008}. In this way, similar profile points are located closer to each other. In this work, it was used as an exploratory data technique to evidence associations between the response category assigned by each panellist ("1 = disliked very or extremely" to "5 = liked extremely or very much") and prebiotic beverage formulations ($F_1$ to $F_{13}$), concerning sensory attributes (1: body, 2: flavour, 3: sweetness, 4: overall impression and 5: aroma). In order to use this method, initially it was verified associations among the prebiotic beverage formulations ($F_1$ to $F_{13}$), the response categories (``1 = I disliked a lot or I disliked it extremely'' to ``5 = I liked it extremely or I liked it very much'') for each the sensory attributes, using a Chi-square test (p-value $< 0.001$).

\begin{figure}[H]
    \centering
    \includegraphics[width=0.85\textwidth]{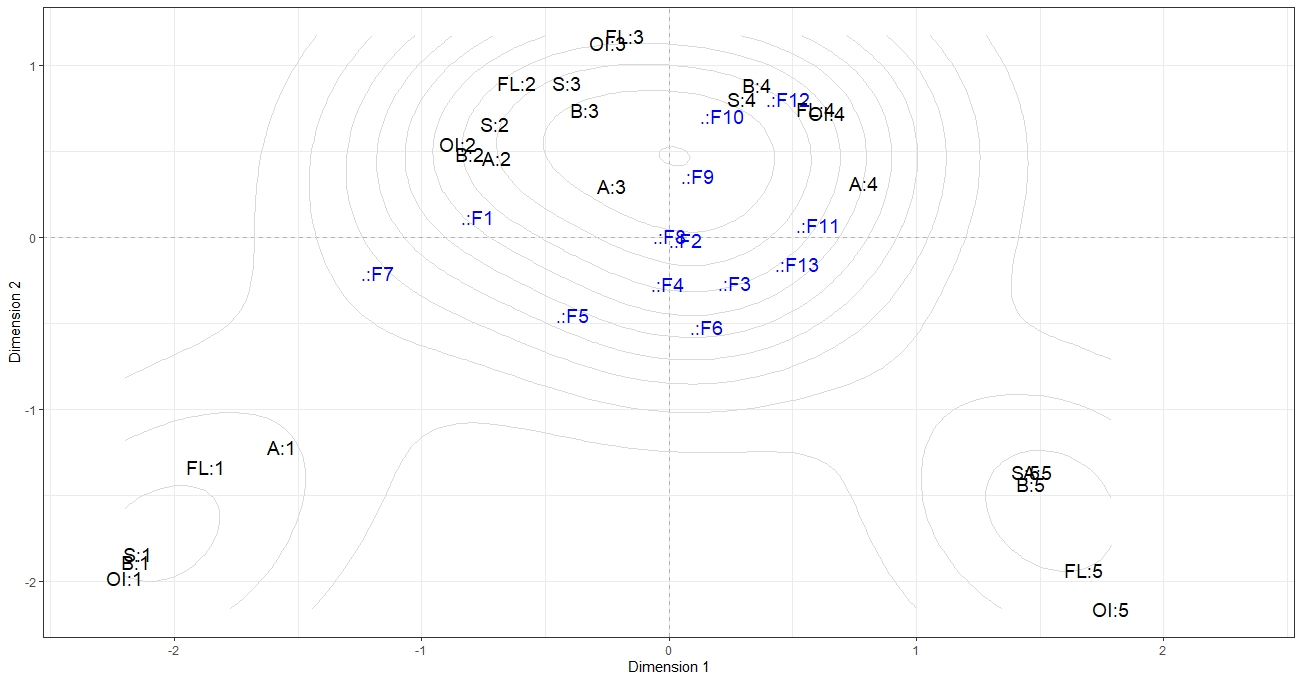}
    \caption{Two-dimensional graphic representation through the analysis of multiple correspondences between prebiotic beverage formulations (F1 to F13), sensory attributes (OI: overall impression, A: aroma, T: body, S: sweetness, FL: flavour) and the categories of responses of the tasters (1 to 5), from the study developed at the Federal University of Ceará, in the year 2016} 
	\label{a2Fig:ACM}
\end{figure}

The distribution of beverage formulations ($F_1$ to $F_{13}$) and sensory attributes (overall impression, aroma, body, sweetness and flavour) was graphically represented by the correspondence analysis as shown in Figure \ref{a2Fig:ACM}.

According to Figure \ref{a2Fig:ACM}  the formulations $F_{3}$ ($4\%$ sugar and $40\%$ grape juice), $F_{6}$ ($6\%$ sugar and $44\%$ grape juice) and $F_{13}$ ($6\%$ sugar and $30\%$ grape juice) were the ones that came closest to response category 5 (liked extremely or very much) with respect to all sensory attributes. The formulations $F_{3}$ and $F_{6}$ have one of the highest concentrations of grape juice (with $40\%$ and $44\%$, respectively), thus indicating that panellists appreciated formulations with higher concentrations of grape juice. On the other hand, the $F_{13}$ formulation represents the midpoint prebiotic beverage, that is, the average percentage of the grape juice ($30\%$) and sugar ($6\%$) levels.

Fitting the unified model, as described in the section (\ref{a2Secao:modeloproposto}), we rejected the constant  proportionality model (p-value $< 0.001$). Thus, it is recommended to use the unified non-proportional odds cumulative logit model, which implies an increase in the number of parameters in the model (\ref{a2eq:modeloproposto}). Moreover, it was found that there is an effect of beverage formulation (p-value $< 0.01$) and sensory attribute (p-value $< 0.01$) by the likelihood ratio test. Thus, these two covariates were considered in the model. To verify if the panellist has a random component, the profiled likelihood ratio test was used as described at the section \ref{a2Secao:modeloproposto}. The estimated standard deviation of random effect was $\hat{\sigma}_{u} = 1.89$ and the 95\% confidence interval for $\sigma^2$ was $IC(\sigma^2)_{95\%} = (1.70; 2.24)$, which does not contain the zero value, confirming the existence of a panelist random effect.Consequently, we have selected the unified mixed model, whose estimated parameters,  standard errors and p-values  are shown in Table \ref{Tab:EstimativadoModelo}.

The second column of the Table \ref{Tab:EstimativadoModelo} represent the parameter estimates, $\hat{\beta}_{jt}$ and $\hat{\delta}_{jl}$, associated with the effects of formulations and attributes, respectively,  $j=2,3,4,5$; $t=2,3, \ldots, 13$ and $l=1,2, \ldots, 5$. Response and formulations were taken as references for the first level of the factors. Also, level 5 for attributes was the reference category. It was observed that the smaller estimated values of $\beta$ were associated with best-rated formulations as $F_4$, $F_6$, and $F_{13}$. On the other hand, the bigger values of $\beta$ were associated with less accepted formulations. The estimated coefficients made it possible to predict the probabilities associated with each response category for each formulation and attribute. For example, to calculate $P( Y \leq 3 \mid F=4, A=1)$ we have used $\hat{\alpha}_{3}=0.04$; $\hat{\beta}_{34}=-1.84$ and $\hat{\delta}_{31}=-0.26$ plus the random effect by means using the equation \ref{Probab_mixed}, consequently $P( Y \geq 4 \mid F=4, A=1) = P( Y = 4 \mid F=4, A=1)1 + P( Y = 5 \mid F=4, A=1)= 1 - P( Y \leq 3 \mid F=4, A=1)$.  Similarly, all other associated probabilities can be calculated.

\begin{table}[H]
    \centering
    \scriptsize 
    \caption{Estimated parameters, standard errors and p-values of the unified mixed non-proportional odds cumulative logit model for the analysis of sensory attributes referring to the study developed at the Federal University of Ceará, in 2016 \label{Tab:EstimativadoModelo}} 
    \begin{adjustbox}{max width=\textwidth}
         \begin{tabular}{cccr|cccr}
            \hline
            Parameter & Estimat. & S.E. & $p$-value & Parameter & Estimat. & S.E. & $p$-value \\
            \hline
             $\alpha_2$ & -1.92 & 0.31 & $<$ 0.01 & $\alpha_4$ & 2.02 & 0.27 & $<$ 0.01 \\ 
            $\alpha_3$ & 0.04  & 0.27 & 0.87     & $\alpha_5$ & 4.29 & 0.38 & $<$ 0.01 \\ 
            \hline
            $\beta_{22}$ & -1.36 & 0.34 & $<$ 0.01 & $\beta_{210}$ & -2.28 & 0.45 & $<$ 0.01 \\ 
            $\beta_{32}$ & -0.84 & 0.25 & $<$ 0.01 & $\beta_{310}$ & -1.16 & 0.27 & $<$ 0.01 \\ 
            $\beta_{42}$ & -1.08 & 0.26 & $<$ 0.01 & $\beta_{410}$ & -1.26 & 0.26 & $<$ 0.01 \\ 
            $\beta_{52}$ & -0.97 & 0.37 & 0.01     & $\beta_{510}$ & -0.87 & 0.39 & 0.03 \\ 
            $\beta_{23}$ & -0.61 & 0.35 & 0.08     & $\beta_{211}$ & -1.93 & 0.42 & $<$ 0.01 \\ 
            $\beta_{33}$ & -0.75 & 0.27 & 0.01     & $\beta_{311}$ & -1.65 & 0.29 & $<$ 0.01 \\ 
            $\beta_{43}$ & -1.05 & 0.27 & $<$ 0.01 & $\beta_{411}$ & -1.75 & 0.27 & $<$ 0.01 \\ 
            $\beta_{53}$ & -0.94 & 0.37 & 0.01     & $\beta_{511}$ & -1.17 & 0.38 & $<$ 0.01 \\ 
            $\beta_{24}$ & -1.22 & 0.32 & $<$ 0.01 & $\beta_{212}$ & -3.28 & 0.71 & $<$ 0.01 \\ 
            $\beta_{34}$ & -1.84 & 0.27 & $<$ 0.01 & $\beta_{312}$ & -1.80 & 0.31 & $<$ 0.01 \\ 
            $\beta_{44}$ & -1.94 & 0.27 & $<$ 0.01 & $\beta_{412}$ & -1.88 & 0.29 & $<$ 0.01 \\ 
            $\beta_{54}$ & -1.66 & 0.39 & $<$ 0.01 & $\beta_{512}$ & -0.91 & 0.42 & 0.03 \\ 
            $\beta_{25}$ & -0.50 & 0.31 & 0.11     & $\beta_{213}$ & -2.31 & 0.42 & $<$ 0.01 \\ 
            $\beta_{35}$ & -0.92 & 0.27 & $<$ 0.01 & $\beta_{313}$ & -2.00 & 0.29 & $<$ 0.01 \\ 
            $\beta_{45}$ & -0.99 & 0.27 & $<$ 0.01 & $\beta_{413}$ & -2.09 & 0.27 & $<$ 0.01 \\ 
            $\beta_{55}$ & -0.87 & 0.40 & 0.03     & $\beta_{513}$ & -1.91 & 0.37 & $<$ 0.01 \\ 
            $\beta_{26}$ & -1.78 & 0.34 & $<$ 0.01 & $\delta_{21}$ & -0.05 & 0.23 & 0.84 \\ 
            $\beta_{36}$ & -1.96 & 0.27 & $<$ 0.01 & $\delta_{31}$ & -0.26 & 0.16 & 0.11 \\ 
            $\beta_{46}$ & -2.25 & 0.27 & $<$ 0.01 & $\delta_{41}$ &  0.81 & 0.16 & $<$ 0.01 \\ 
            $\beta_{56}$ & -2.44 & 0.38 & $<$ 0.01 & $\delta_{51}$ &  0.35 & 0.22 & 0.12 \\ 
            $\beta_{27}$ & -0.21 & 0.30 & 0.47     & $\delta_{22}$ & -0.12 & 0.23 & 0.60 \\ 
            $\beta_{37}$ &  0.03 & 0.26 & 0.92     & $\delta_{32}$ & -0.41 & 0.16 & 0.01 \\ 
            $\beta_{47}$ &  0.24 & 0.28 & 0.38     & $\delta_{42}$ & -0.38 & 0.15 & 0.01 \\ 
            $\beta_{57}$ & -0.61 & 0.44 & 0.17     & $\delta_{52}$ & -0.72 & 0.20 & $<$ 0.01 \\ 
            $\beta_{28}$ & -1.75 & 0.35 & $<$ 0.01 & $\delta_{23}$ &  0.15 & 0.23 & 0.51 \\
            $\beta_{38}$ & -1.58 & 0.27 & $<$ 0.01 & $\delta_{33}$ & -0.42 & 0.16 & 0.01 \\ 
            $\beta_{48}$ & -1.84 & 0.27 & $<$ 0.01 & $\delta_{43}$ & -0.52 & 0.16 & $<$ 0.01 \\
            $\beta_{58}$ & -1.23 & 0.40 & $<$ 0.01 & $\delta_{53}$ & -0.84 & 0.20 & $<$ 0.01 \\
            $\beta_{29}$ & -1.60 & 0.36 & $<$ 0.01 & $\delta_{24}$ &  0.81 & 0.21 & $<$ 0.01 \\ 
            $\beta_{39}$ & -1.45 & 0.27 & $<$ 0.01 & $\delta_{34}$ &  0.23 & 0.16 & 0.14 \\ 
            $\beta_{49}$ & -1.65 & 0.27 & $<$ 0.01 & $\delta_{44}$ & -0.03 & 0.15 & 0.85 \\ 
            $\beta_{59}$ & -1.27 & 0.38 & $<$ 0.01 & $\delta_{54}$ & -0.19 & 0.21 & 0.35 \\
            \hline
        \end{tabular} 
        \end{adjustbox}
\end{table}

The criterion used to select the prebiotic beverage formulations more accepted, considering all sensory attributes, was the accumulated probability between categories 4 (I liked it moderately or slightly) and 5 (I liked it extremely or very much) predicted according to the fitted mixed model, which is presented in the Table \ref{Tab:TabeladeProporcoesEstimadoAtributos}.

\begin{table}[!htbp]
  \centering
	\caption{Response categories probabilities according to sensory attributes and the predicted accumulated probabilities for the categories 4 and 5, referring to the study developed at the Federal University of Ceará, in 2016}  \label{Tab:TabeladeProporcoesEstimadoAtributos}
    \begin{adjustbox}{max width=\textwidth, max totalheight=0.9\textheight, keepaspectratio}
    \begin{tabular}
{c|c|p{.6cm}p{.6cm}p{.6cm}p{.6cm}p{.6cm}p{.6cm}p{.6cm}p{.6cm}|p{.6cm}p{.6cm}p{.6cm}p{.6cm}p{.6cm}}
    \hline
  \multirow{1}[2]{*}{Sensory} &  \multirow{1}[2]{*}{Response} & \multicolumn{13}{c}{Beverage formulations} \\
\cmidrule{3-15} Attributes & Categories & $F_1$ & $F_2$ & $F_3$ & \cellcolor{gray!10} $F_4$ & $F_5$ & \cellcolor{gray!10} $F_6$ & $F_7$ &  $F_8$ & $F_9$ & $F_{10}$ & $F_{11}$ & $F_{12}$ & \cellcolor{gray!10} $F_{13}$ \\
    \hline

    & 1 & 0.13  & 0.04  & 0.08  & \cellcolor{gray!10} 0.04  & 0.09  & \cellcolor{gray!10} 0.03  & 0.11  &  0.03  & 0.03  & 0.02  & 0.02  & 0.01  & \cellcolor{gray!10} 0.02 \\
                        & 2 & 0.44  & 0.33  & 0.31  & \cellcolor{gray!10} 0.13  & 0.26  & \cellcolor{gray!10} 0.13  & 0.47  &  0.19  & 0.21  & 0.28  & 0.18  & 0.18  & \cellcolor{gray!10} 0.14 \\
    Aroma                    & 3 & 0.20  & 0.16  & 0.15  & \cellcolor{gray!10} 0.16  & 0.21  & \cellcolor{gray!10} 0.10  & 0.23  &  0.13  & 0.15  & 0.19  & 0.17  & 0.14  & \cellcolor{gray!10} 0.13 \\
                        & 4 & 0.21  & 0.42  & 0.41  & \cellcolor{gray!10} 0.58  & 0.40  & \cellcolor{gray!10} 0.56  & 0.16  &  0.59  & 0.54  & 0.47  & 0.57  & 0.62  & \cellcolor{gray!10} 0.59 \\
                        & 5 & 0.02  & 0.05  & 0.05  & \cellcolor{gray!10} 0.09  & 0.04  & \cellcolor{gray!10} 0.18  & 0.03  &  0.06  & 0.07  & 0.04  & 0.06  & 0.05  & \cellcolor{gray!10} 0.12 \\
\cmidrule{2-15}          
                        & $\mathds{P}(Y \geq 4)$ & 0.23  & 0.47  & 0.46  & \cellcolor{gray!10} 0.67  & 0.44  & \cellcolor{gray!10} 0.74  & 0.19  &  0.65  & 0.61  & 0.51  & 0.63  & 0.67  & \cellcolor{gray!10} 0.71 \\
\hline

          & 1 & 0.13  & 0.04  & 0.07  & \cellcolor{gray!10} 0.04  & 0.08  & \cellcolor{gray!10} 0.02  & 0.11  &  0.02  & 0.03  & 0.01  & 0.02  & 0.01  & \cellcolor{gray!10} 0.01 \\
          & 2 & 0.38  & 0.27  & 0.26  & \cellcolor{gray!10} 0.10  & 0.21  & \cellcolor{gray!10} 0.10  & 0.41  &  0.15  & 0.17  & 0.23  & 0.15  & 0.14  & \cellcolor{gray!10} 0.11 \\
    Body      & 3 & 0.38  & 0.41  & 0.40  & \cellcolor{gray!10} 0.38  & 0.45  & \cellcolor{gray!10} 0.32  & 0.39  & 0.37  & 0.39  & 0.44  & 0.40  & 0.39  & \cellcolor{gray!10} 0.36 \\
          & 4 & 0.10  & 0.25  & 0.24  & \cellcolor{gray!10} 0.41  & 0.23  & \cellcolor{gray!10} 0.42  & 0.07  & 0.41  & 0.36  & 0.29  & 0.39  & 0.43  & \cellcolor{gray!10} 0.43 \\
          & 5 & 0.01  & 0.03  & 0.03  & \cellcolor{gray!10} 0.07  & 0.03  & \cellcolor{gray!10} 0.14  & 0.02  &  0.05  & 0.05  & 0.03  & 0.04  & 0.03  & \cellcolor{gray!10} 0.09 \\
\cmidrule{2-15}          
         & $\mathds{P}(Y \geq 4)$ & 0.11  & 0.28  & 0.27  & \cellcolor{gray!10} 0.48  & 0.26  & \cellcolor{gray!10} 0.56  & 0.09  &  0.46  & 0.41  & 0.32  & 0.43  & 0.46  & \cellcolor{gray!10} 0.52 \\
    \hline
    
          & 1 & 0.15  & 0.04  & 0.09  & \cellcolor{gray!10} 0.05  & 0.10  & \cellcolor{gray!10} 0.03  & 0.13  &  0.03  & 0.03  & 0.02  & 0.03  & 0.01  & \cellcolor{gray!10} 0.02 \\
          & 2 & 0.32  & 0.23  & 0.21  & \cellcolor{gray!10} 0.07  & 0.16  & \cellcolor{gray!10} 0.08  & 0.35  & 0.12  & 0.14  & 0.20  & 0.12  & 0.12  & \cellcolor{gray!10} 0.09 \\
    Sweetness      & 3 & 0.20  & 0.13  & 0.11  & \cellcolor{gray!10} 0.10  & 0.16  & \cellcolor{gray!10} 0.06  & 0.23  &  0.09  & 0.10  & 0.14  & 0.10  & 0.10  & \cellcolor{gray!10} 0.09 \\
          & 4 & 0.29  & 0.49  & 0.49  & \cellcolor{gray!10} 0.59  & 0.48  & \cellcolor{gray!10} 0.49  & 0.21  &  0.63  & 0.59  & 0.54  & 0.62  & 0.67  & \cellcolor{gray!10} 0.57 \\
          & 5 & 0.04  & 0.11  & 0.10  & \cellcolor{gray!10} 0.19  & 0.10  & \cellcolor{gray!10} 0.34  & 0.08  &  0.13  & 0.14  & 0.10  & 0.13  & 0.10  & \cellcolor{gray!10} 0.23 \\
\cmidrule{2-15}          & $\mathds{P}(Y \geq 4)$ & 0.33  & 0.60  & 0.59  & \cellcolor{gray!10} 0.78  & 0.58  & \cellcolor{gray!10} 0.83  & 0.29  &  0.76  & 0.73  & 0.64  & 0.75  & 0.77  & \cellcolor{gray!10} 0.80 \\
    \hline
    
          & 1 & 0.12  & 0.03  & 0.07  & \cellcolor{gray!10} 0.04  & 0.08  & \cellcolor{gray!10} 0.02  & 0.10  &  0.02  & 0.03  & 0.01  & 0.02  & 0.01  & \cellcolor{gray!10} 0.01 \\
          & 2 & 0.35  & 0.25  & 0.23  & \cellcolor{gray!10} 0.09  & 0.19  & \cellcolor{gray!10} 0.09  & 0.38  &  0.13  & 0.15  & 0.21  & 0.13  & 0.12  & \cellcolor{gray!10} 0.10 \\
    Flavor      & 3 & 0.22  & 0.15  & 0.15  & \cellcolor{gray!10} 0.12  & 0.19  & \cellcolor{gray!10} 0.08  & 0.28  &  0.12  & 0.13  & 0.17  & 0.14  & 0.13  & \cellcolor{gray!10} 0.11 \\
          & 4 & 0.27  & 0.47  & 0.46  & \cellcolor{gray!10} 0.58  & 0.45  & \cellcolor{gray!10} 0.49  & 0.19  &  0.61  & 0.57  & 0.52  & 0.60  & 0.65  & \cellcolor{gray!10} 0.57 \\
          & 5 & 0.04  & 0.10  & 0.09  & \cellcolor{gray!10} 0.17  & 0.09  & \cellcolor{gray!10} 0.31  & 0.07  &  0.12  & 0.12  & 0.09  & 0.11  & 0.09  & \cellcolor{gray!10} 0.21 \\
\cmidrule{2-15}          & $\mathds{P}(Y \geq 4)$ & 0.31  & 0.57  & 0.55  & \cellcolor{gray!10} 0.75  & 0.54  & \cellcolor{gray!10} 0.80  & 0.26  &  0.73  & 0.69  & 0.61  & 0.71  & 0.74  & \cellcolor{gray!10} 0.78 \\
    \hline

          & 1 & 0.26  & 0.08  & 0.16  & \cellcolor{gray!10} 0.09  & 0.17  & \cellcolor{gray!10} 0.06  & 0.22  &  0.06  & 0.07  & 0.03  & 0.05  & 0.01  & \cellcolor{gray!10} 0.03 \\
          & 2 & 0.37  & 0.34  & 0.29  & \cellcolor{gray!10} 0.12  & 0.23  & \cellcolor{gray!10} 0.14  & 0.42  &  0.20  & 0.22  & 0.31  & 0.20  & 0.21  & \cellcolor{gray!10} 0.15 \\
OI          & 3 & 0.14  & 0.10  & 0.08  & \cellcolor{gray!10} 0.11  & 0.15  & \cellcolor{gray!10} 0.06  & 0.17  &  0.08  & 0.09  & 0.14  & 0.11  & 0.11  & \cellcolor{gray!10} 0.10 \\
          & 4 & 0.21  & 0.42  & 0.41  & \cellcolor{gray!10} 0.57  & 0.40  & \cellcolor{gray!10} 0.53  & 0.15  &  0.59  & 0.54  & 0.47  & 0.57  & 0.61  & \cellcolor{gray!10} 0.58 \\
          & 5 & 0.02  & 0.06  & 0.06  & \cellcolor{gray!10} 0.11  & 0.05  & \cellcolor{gray!10} 0.21  & 0.04  &  0.07  & 0.08  & 0.05  & 0.07  & 0.06  & \cellcolor{gray!10} 0.14 \\
\cmidrule{2-15}          & $\mathds{P}(Y \geq 4)$ & 0.23  & 0.48  & 0.47  & \cellcolor{gray!10} 0.68  & 0.45  & \cellcolor{gray!10} 0.74  & 0.19  &  0.66  & 0.62  & 0.52  & 0.64  & 0.67  & \cellcolor{gray!10} 0.72 \\
    \hline
    \end{tabular}%
    \end{adjustbox}
\end{table}%

In Table \ref{Tab:TabeladeProporcoesEstimadoAtributos}, it is observed that the central point prebiotic formulations ($F_9$ to $F_{13}$, with 6\% of sugar and 30\% of grape juice) have the accumulated probabilities (4 and 5 categories) varying between  0.32 to 0.80 considering all sensory attributes. It is worth mentioning that the formulation $F_{13}$ (central point) stands out more among the others for all attributes whose accumulated probability are 0.71 for aroma, 0.52 for the body, 0.80 for sweetness, 0.78 for flavour and 0.72 for overall impression, respectively.

\begin{figure}[H]
    \centering
    \includegraphics[scale=.5]{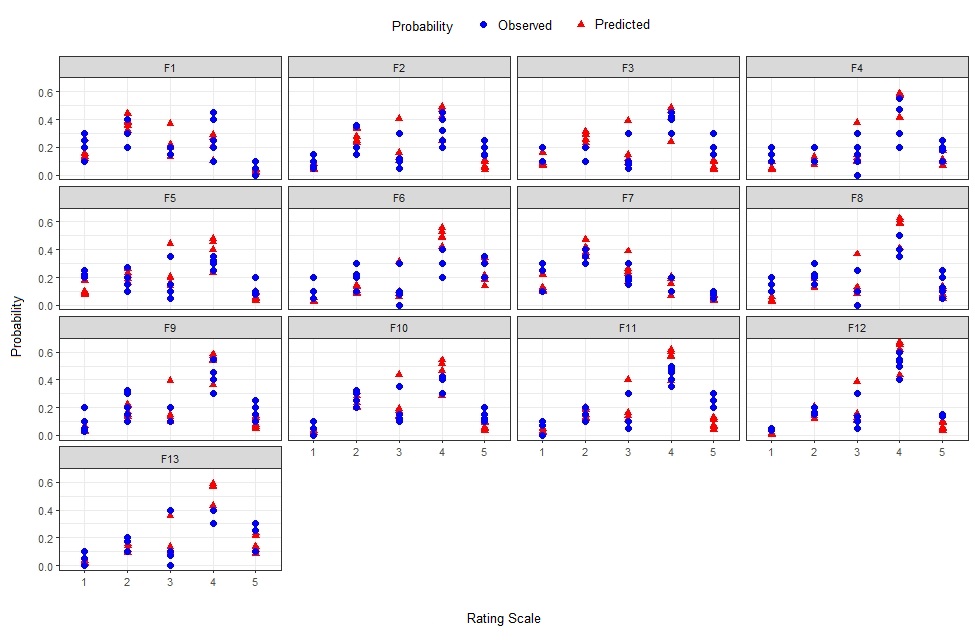}
    \caption{Observed and predicted probabilities for prebiotic beverage formulations in relation to  five sensory attributes referring to the study developed at the Federal University of Ceará, in 2016} 
	\label{Fig:PrevistoObservadoMult}
\end{figure}

Using this same criterion, the other beverage formulations,  $F_4$ (8\% sugar and 40\% grape juice) and $F_6$ (6\% sugar and 44\% of grape juice), have the accumulated probabilities, respectively: 0.67 and 0.74 for aroma;  0.48 and 0.56 for the body;  0.78 and 0.83 for sweetness; 0.75 and 0.80 for flavour and 0.68 and 0.74 for overall impression. It is observed that the most accepted formulation concerning all sensory attributes is composed of 6\% sugar and 44\% grape juice ($F_6$), and the least accepted was $F_7$ with 3\% of sugar and 30\% grape juice. The observed and predicted probabilities by the model are close, with rare exceptions, indicating that the unified mixed non-proportional odds model has a good performance, as shown in the Figure (\ref{Fig:PrevistoObservadoMult}).

As a result, the formulations with the highest acceptability through the proposed unified model were $F_4$ (8\% sugar and 40\% grape juice), $F_6$ (6\% sugar and 44\% grape juice) and $F_{13}$ (6\% sugar and 30\% grape juice), which could give greater emphasis to the formulation $F_6$, whose the beverage concentration grape juice was 44\% and a level sugar intermediate of 6\%. In addition, it was identified that the three most accepted prebiotic beverage formulations were in the first quadrant of the $2^2$ factorial rotational central composite design. That is, the selected formulations are the ones that had the highest concentrations of grape juice (with 30\%, 40\% and 44\%) and sugar (with 6\% and 8\%), thus indicating a preference for sweeter juices with greater concentrations in fruit juice.

\section{Conclusion}

Generalized cumulative logit models, proportional or not, are an important framework for data analysis with hedonic scales ~(categorical and ordinal), especially in sensory studies. Although the analysis of the design's sensory attributes can be done separately, the unified model allows a singular selection of the best products. Furthermore, when separate models are fitted to data for each attribute, an odds-proportional condition may occur for some models and failure for others, making diagnostic techniques difficult. However, we have the advantage of making a unique analysis. In this way, the model selection, choice of formulations, predictions, and residual analysis are done simultaneously in a parsimonious way. Regarding the study of motivation, the results obtained by the single analysis showed that the best formulations were $F_{4}$, $F_{6}$ and $F_{13}$, results compatible with the separate evaluation of each attribute. 

Our simulation studies, even in more restricted scenarios, showed good performance of the unified model, simplifying the process of adjusting and selecting formulations. It is noted that this method is an introductory approach to unified sensory analysis. In studies with the full hedonic scale, with 9 points, problems of excessive parameterization may arise, but this is a common problem with categorical data. Another restrictive point of our proposal is the excess of parameters, especially if the proportionality condition fails, as in the illustrative example. Therefore, future simulation studies are needed to improve this method, including residual analysis. Despite the excess of parameters, in the applied area, choosing the product based on some criteria is important. In this sense, the method presented is also an important decision-making tool, especially for sensometry.




\section*{Acknowledgments}
\justifying
This work was supported by funding: Coordenação de Aperfeiçoamento de Pessoal de Nível Superior~(CAPES), process numbers (88882.378345/2019-01) and (88887.821274/2023-00), and Taighde Éireann – Research Ireland under Grant number 18/CRT/6049.

\subsection*{Conflicts of Interest}
The authors declare that there is no conflict of interest in this work.

\bibliographystyle{apalike}
\bibliography{ref}

\end{document}